\newif\ifarxiv
\arxivfalse 
\ifarxiv
\documentclass{article}
\usepackage{arxiv}
\else
\documentclass[conference]{IEEEtran}
\IEEEoverridecommandlockouts
\fi

\pdfoutput=1

\usepackage{cite}
\usepackage{amsmath,amssymb,amsfonts}
\usepackage{algorithm}
\usepackage{algpseudocode}
\usepackage{graphicx}
\usepackage{textcomp}
\usepackage{xcolor}
\usepackage{subcaption}
\usepackage{multirow}

\DeclareMathOperator*{\argmax}{arg\,max}

\def\BibTeX{{\rm B\kern-.05em{\sc i\kern-.025em b}\kern-.08em
    T\kern-.1667em\lower.7ex\hbox{E}\kern-.125emX}}

\begin{document}

\title{Generalized compression and compressive search of large datasets}

\ifarxiv
\else

\author{\IEEEauthorblockN{Morgan~E.~Prior}
\IEEEauthorblockA{\textit{Dept. of Computer Science} \\
\textit{Tufts University}\\
Boston, MA, USA \\
morgan.prior@tufts.edu \\
0009-0003-1553-1672}
\and
\IEEEauthorblockN{Thomas~J.~Howard~III}
\IEEEauthorblockA{\textit{Dept. of Computer Science and Statistics} \\
\textit{University of Rhode Island}\\
Kingston, RI, USA\\
thoward27@uri.edu \\
0000-0001-6051-2508}
\and
\IEEEauthorblockN{Emily~Light}
\IEEEauthorblockA{\textit{Dept. of Computer Science and Statistics} \\
\textit{University of Rhode Island}\\
Kingston, RI, USA\\
emily\_light@uri.edu \\
}
\and
\IEEEauthorblockN{Najib~Ishaq}
\IEEEauthorblockA{\textit{Dept. of Computer Science and Statistics} \\
\textit{University of Rhode Island}\\
Kingston, RI, USA\\
najib\_ishaq@zoho.com \\
0009-0009-0830-0124}
\and
\IEEEauthorblockN{Noah~M.~Daniels}
\IEEEauthorblockA{\textit{Dept. of Computer Science and Statistics} \\
\textit{University of Rhode Island}\\
Kingston, RI, USA\\
noah\_daniels@uri.edu \\
0000-0002-9538-825X}
}

\fi

\maketitle

\begin{abstract}
    The Big Data explosion has necessitated the development of search algorithms that scale sub-linearly in time and memory.
    While compression algorithms and search algorithms do exist independently, few algorithms offer both, and those which do are domain-specific.
    We present panCAKES, a novel approach to compressive search, i.e., a way to perform $k$-NN and $\rho$-NN search on compressed data while only decompressing a small, relevant, portion of the data.
    panCAKES assumes the manifold hypothesis and leverages the low-dimensional structure of the data to compress and search it efficiently.
    panCAKES is generic over any distance function for which the distance between two points is proportional to the memory cost of storing an encoding of one in terms of the other.
    This property holds for many widely-used distance functions, e.g. string edit distances (Levenshtein, Needleman-Wunsch, etc.) and set dissimilarity measures (Jaccard, Dice, etc.).
    We benchmark panCAKES on a variety of datasets, including genomic, proteomic, and set data.
    We compare compression ratios to gzip, and search performance between the compressed and uncompressed versions of the same dataset.
    panCAKES achieves compression ratios close to those of gzip, while offering sub-linear time performance for $k$-NN and $\rho$-NN search.
    We conclude that panCAKES is an efficient, general-purpose algorithm for exact compressive search on large datasets that obey the manifold hypothesis.
    We provide an open-source implementation of panCAKES in the Rust programming language.
\end{abstract}

\ifarxiv
\else
\fi

\section{Introduction}
\label{sec:introduction}

Researchers are collecting data at an unprecedented scale. With the cardinalities and dimensionalities of datasets growing exponentially, scientists in myriad disciplines need new algorithms to address this big data deluge.

In the field of genomics, for example, improvements in sequencing methods allow for collection of massive biological datasets.
Examples of such datasets include the GreenGenes project~\cite{desantis2006greengenes}, which provides a multiple-sequence alignment of over one million bacterial 16S sequences each 7,682 characters in length, and SILVA 18S~\cite{quast2012silva}, which contains ribosomal DNA sequences of approximately 2.25 million genomes with an aligned length of 50,000 letters.
With these large quantities of high dimensional data, storage and computational costs have replaced sequencing as the primary research bottleneck~\cite{berger2022bottlenecks}.

Other fields, especially those which use datasets of set-membership vectors, face different big data challenges.
Given that these datasets often have dimensionality on the same order of magnitude as cardinality, the data only sparsely populate the space.
Algorithms which do not leverage the sparsity and self-similarity in these datasets are often prohibitively slow~\cite{prior2023cakes}.

As the sizes of datasets grow, the ability to compress them has become increasingly valuable.
However, researchers who wish to perform analysis on their compressed data in the near-term face an additional challenge: the computational cost of decompressing the data before analysis.
One particular type of analysis used on large datasets in many fields is \emph{similarity} search.
Similarity search enables a variety of applications, including classification systems~\cite{suyanto2022knnclassifier} and genetic sequence analysis~\cite{frith2020simsearch}.

As described in \cite{prior2023cakes}, there are two common definitions of similarity search: $k$-nearest neighbor search ($k$-NN) and $\rho$-nearest neighbor search ($\rho$-NN).
Given some measure of similarity between data points (e.g., a distance function), $k$-NN search aims to find the $k$ most similar points to a query, while $\rho$-NN search aims to find all points within a similarity threshold $\rho$ of a query.

Similarity search algorithms often exhibit steep tradeoffs between recall and throughput~\cite{Malkov2016EfficientAR, johnson2019billion, annoy, aumuller2020ann}.
Previous works have used the term \textit{approximate} search to refer to $\rho$-NN search, but in this paper, as in \cite{prior2023cakes}, we reserve the term \textit{approximate} for search algorithms which do not exhibit perfect recall when compared to a na\"{i}ve linear search. In contrast, an \textit{exact} search algorithm exhibits perfect recall.

As mentioned above, speed versus accuracy is not the only tradeoff researchers performing similarity search must grapple with.
Those working with compressed data also face the time and space tradeoff of decompressing data before performing search;
often, the better the compression ratio, the more computationally intensive the decompression process.

\emph{Compressive search} algorithms aim to improve this space-time tradeoff.
We define compressive search to be the problem of performing search on a compressed dataset without decompressing the entire dataset.
Ideally, the only points we decompress are those in the results set, but in practice, we must decompress additional points in order to identify the correct results set.

Few compressive search algorithms exist~\cite{berger2022bottlenecks}.
Though the ``compressive genomics'' technique presented in~\cite{daniels2013compressive} has proven effective, the approach is specially tailored to genomics data and does not generalize to other domains.
Existing general-purpose compression algorithms like gzip~\cite{gzip} require decompression of the \emph{entire} dataset before performing similarity search.

The compressive search algorithm presented in this paper is general over any distance function $f$ which, for all points $x$ and $y$, satisfies the following properties:

\begin{itemize}
    \item $f$ provides an ``encoding'' of $x$ in terms of $y$ and vice-versa, i.e. a way to transform $x$ into $y$ and $y$ into $x$, and
    \item the memory required to store the ``encoding'' is proportional to the distance between $x$ and $y$.
\end{itemize}

In \cite{ishaq2019clustered}, we introduced a novel hierarchical clustering algorithm, and in \cite{prior2023cakes}, we introduced novel algorithms for efficient, exact $k$-NN search on large datasets. Both of these approaches rely on the construction of a cluster tree, called the CLAM tree. 
In this paper, we extend these previous works by introducing a novel hierarchical compression algorithm, panCAKES (pan- CLAM-Accelerated K-NN Entropy Scaling Search), which uses the indexing approach from \cite{ishaq2019clustered} and \cite{prior2023cakes} to enable efficient, exact $k$-NN and $\rho$-NN search on compressed data.
We stress that panCAKES' compression is aligned with, rather than orthogonal to, speedups provided by our indexing approach.
Our hierarchical cluster tree allows for both efficient compression and search on the compressed dataset. Importantly, panCAKES conducts search on compressed datasets without needing modifications to the $k$-NN and $\rho$-NN search algorithms from~\cite{prior2023cakes}.

The motivation behind panCAKES is that many interesting datasets are too large to fit in system memory, and therefore searching them requires significant I/O. In contrast, the compressive search provided by panCAKES is designed to search datasets that do not fit in memory, while decompressing only relevant subsets of the data. 

We benchmark panCAKES on two genomic datasets, SILVA-18S~\cite{quast2012silva} and GreenGenes~\cite{desantis2006greengenes}, a proteomic dataset, PDB-seq~\cite{bank1971protein}, and two set datasets, Kosarak~\cite{Bodon2003AFA} and MovieLens-10M~\cite{maxwell2015k}.
Additionally, we compare our compression ratios to those of gzip.
We also provide a comparison between the time taken to perform similarity search on compressed data using panCAKES versus the time taken to run similarity search with algorithms from our previous work in \cite{prior2023cakes}.
panCAKES is implemented in the Rust programming language and the source code is available under an MIT license at https://github.com/URI-ABD/clam.

\section{Methods}
\label{sec:methods}

\subsection{Building the CLAM Tree}
\label{sec:methods:building-the-tree}

The compression approach presented in this paper relies on the divisive hierarchical clustering algorithm introduced in \cite{prior2023cakes} and described in Algorithm~\ref{alg:methods:partition}, which constructs the CLAM tree. 

\begin{algorithm} 
\caption{Partition($C$, $criteria$)} 
\label{alg:methods:partition} 
\begin{algorithmic}[0] 
    \Require $C$, a cluster
    \Require $criteria$, user-specified stopping criteria

    \State $seeds \Leftarrow$ random sample of $\left\lceil \sqrt{|C|} \right\rceil$ points from $C$
    \State $c \Leftarrow$ geometric median of $seeds$
    \State $l \Leftarrow \argmax f(c, x) \ \forall \ x \in C$
    \State $r \Leftarrow \argmax f(l, x) \ \forall \ x \in C$
    \State $L \Leftarrow \{x \ | \ x \in C \land f(l, x) \le f(r, x)\}$
    \State $R \Leftarrow \{x \ | \ x \in C \land f(r, x) < f(l, x)\}$

    \If{$|L| > 1$ \textbf{and} $L$ satisfies $criteria$}
        \State Partition($L$, $criteria$)
    \EndIf

    \If{$|R| > 1$ \textbf{and} $R$ satisfies $criteria$}
        \State Partition($R$, $criteria$)
    \EndIf
\end{algorithmic}
\end{algorithm}

Note that Algorithm~\ref{alg:methods:partition} does not necessarily produce a balanced tree.
In fact, if we assume the \emph{manifold hypothesis}--the notion that high-dimensional data collected from constrained generating phenomena typically only occupy a low-dimensional manifold within their embedding space--we expect anything but a balanced tree.
The varying sampling density in different regions of the manifold and the low dimensional ``shape'' of the manifold itself will cause the tree to be unbalanced.
The only case in which we would expect a balanced tree is if the dataset were uniformly distributed, e.g. in a $d$-dimensional hyper-sphere.
We discuss additional ways that dataset structure informs properties of the cluster tree in Section~\ref{subsec:results:scaling-behavior-of-cluster-radii}.

\subsection{Compression}
\label{sec:methods:compression}

We first describe a compression approach we call \emph{unitary compression} wherein a cluster's points are represented in terms of the cluster's center.
In this approach, given a cluster $C$ with center $c$, for each non-center point $x \in C \setminus c$, we store simply the ``encoding'' required to transform $c$ into $x$.
For example, with a genomic dataset under Levenshtein distance~\cite{levenshtein1966binary}, this encoding would be the sequence of edits (i.e., insertions, deletions, and substitutions) needed to transform $x$ into $c$.
We determine this sequence of edits using the Needleman-Wunsch algorithm~\cite{needleman1970general}.
Or, with a set dataset under Jaccard~\cite{Jaccard1912Jaccard} distance, the encoding would be set differences (i.e., those elements in $x$ but not in the center, and vice versa).
For the distance functions we use, the cost of storing an encoding of point $x$ in terms of center $c$ is proportional to $d(x, c)$, the distance between $x$ and $c$.
Intuitively, when the radius of the cluster is small, or when a large fraction of the cluster's points are close to the center, the cost of unitary compression should be low.

Of course, for massive datasets, radii at shallow depths of the cluster tree are not small, and as discussed in Section~\ref{subsec:results:scaling-behavior-of-cluster-radii}, they are not guaranteed to decrease after the first partition.
Thus, unitary compression may not be the most efficient compression approach for all clusters.

For any non-leaf cluster, we also consider the possibility of compressing it \emph{recursively}.
Let $C$ again denote the cluster we wish to compress, $c$ its center, $L$ and $R$ respective children, and $\ell$ and $r$ the respective child centers.
In this approach, we store the encodings for $\ell$ and $r$ in terms of $c$, along with the compressed form (unitary or recursive, whichever has a lower memory cost) of $L$ and $R$.

\begin{figure}[ht!]
    \centering
    \includegraphics[width=3.4in]{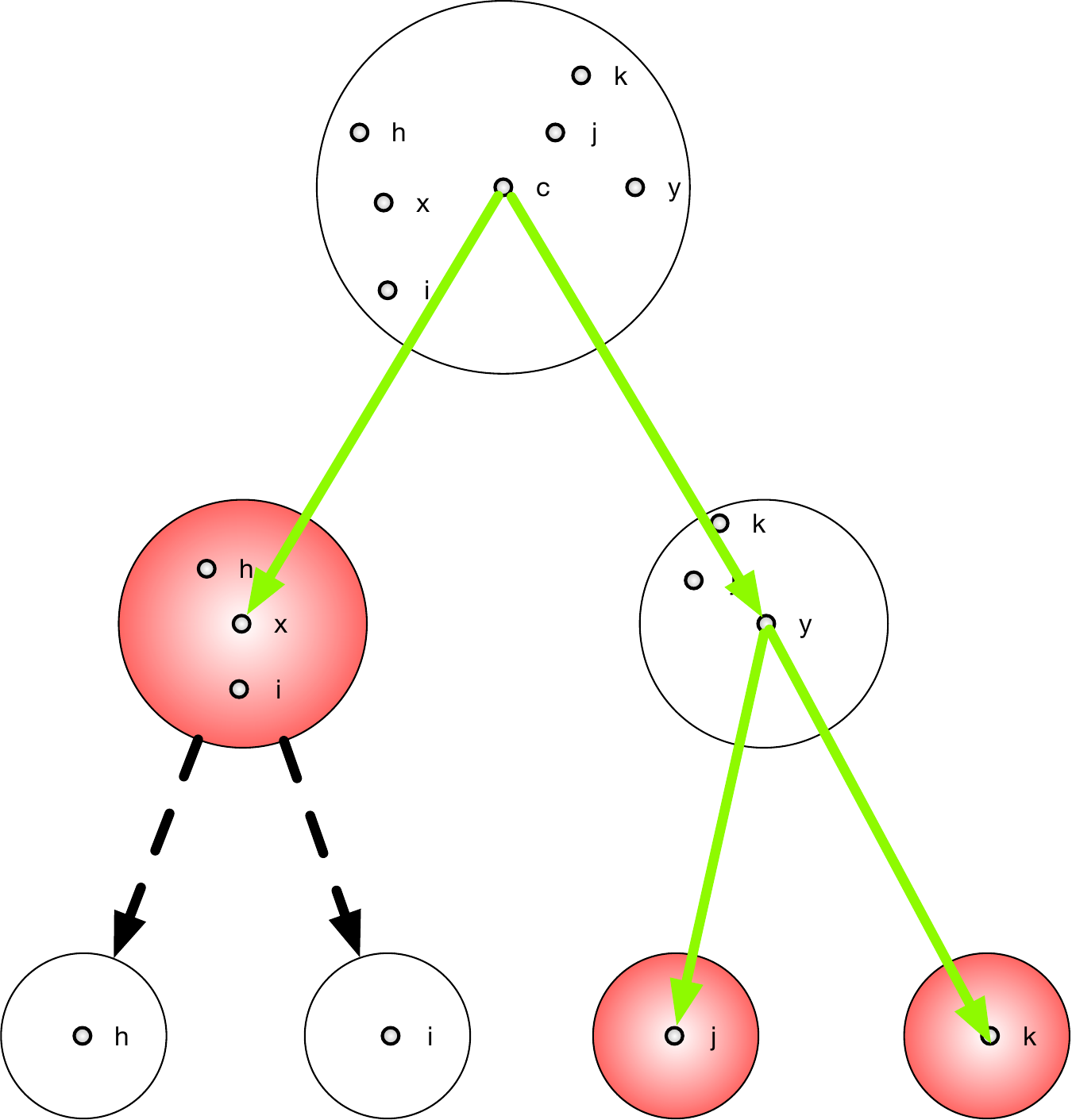}
    \caption{
        A cluster tree with a mix of unitarily and recursively compressed clusters.
        A green edge from a parent to a child indicates that the center of the child will be encoded in terms of the center of the parent.
        A red-shaded cluster indicates that the cluster will be unitarily compressed.
        A dashed edge between a unitarily compressed cluster and a child indicates that the child will be deleted during Algorithm~\ref{alg:methods:compress}.
        Notably, unitarily compressed clusters do not all occur at the same depth in the tree.
        The exact depth at which recursive compression becomes more efficient than unitary compression varies with the structure at different regions of the manifold.
    }
    \label{fig:results:unitary1}
\end{figure}

\begin{figure}[ht!]
    \centering
    \includegraphics[width=3.4in]{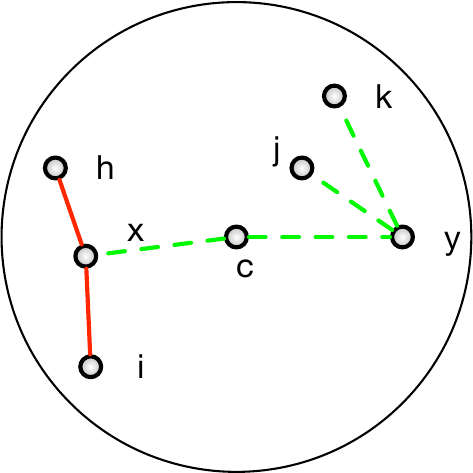}
    \caption{
        An alternate  view of the cluster tree from Figure~\ref{fig:results:unitary1}.
        A green dashed edge between points indicates recursive compression.
        For example, the dashed green edges $\overline{yj}$ and $\overline{yk}$ indicate that $j$ and $k$ were recursively encoded in terms of $y$.
        A solid red edge between points indicates unitary compression.
        For example, the red edge $\overline{xi}$ indicates that $i$ is encoded in terms of its cluster center, $x$.
    }
    \label{fig:results:unitary2}
\end{figure}

We traverse the tree from the root to the leaves, calculating the cost of unitary compression for each cluster.
Then, on the way back up, we also calculate the cost of recursive compression.
For any cluster whose recursive cost is greater than the unitary cost, we delete its descendants and turn it into a leaf.
This procedure is outlined in Algorithm~\ref{alg:methods:compress}.
At the conclusion of this process, we have a compressed tree that may include a mix of unitarily and recursively compressed clusters, as shown in Figures~\ref{fig:results:unitary1} and~\ref{fig:results:unitary2}. 

\begin{algorithm} 
    \caption{Compress($C$)} 
    \label{alg:methods:compress} 
    \begin{algorithmic}[0] 
        \Require $C$, a cluster
        \Require $f$, a distance metric
        \State $C.unitary\_cost \Leftarrow \sum_{x \in C} f(C.center, x)$
        \State $C.min\_cost \Leftarrow C.unitary\_cost$
        \If{$C$ is not a leaf}
            \State $L, R \Leftarrow C.left\_child, C.right\_child$
            \State $\ell, r \Leftarrow L.center, R.center$
            \State Compress($L$)
            \State Compress($R$)
            \State $l\_cost \Leftarrow f(C.center, \ell) + L.min\_cost$
            \State $r\_cost \Leftarrow f(C.center, r) + R.min\_cost$
            \State $C.recursive\_cost \Leftarrow l\_cost + r\_cost$
            \If {$C.recursive\_cost > C.unitary\_cost$}
                \State Delete all descendants of $C$
            \Else
                \State $C.min\_cost \Leftarrow C.recursive\_cost$
            \EndIf
        \EndIf
    \end{algorithmic}
\end{algorithm}

\subsection{Search}
\label{sec:methods:search}

We perform compressed search using the algorithms introduced in ~\cite{prior2023cakes}.
We provide a very brief summary of each algorithm below.
The first algorithm performs $\rho$-NN search, while the other three perform $k$-NN search.

\begin{itemize}
    \item \textbf{$\rho$-NN Search}: Performs search in two stages: tree search and leaf search. During tree search, we determine clusters which overlap the query ball. In leaf search, we linearly scan the points in the overlapping clusters in order to find only those within a distance $\rho$ of the query.
    \item \textbf{Repeated $\rho$-NN Search}: Repeatedly performs $\rho$-NN search with increasing values of $\rho$ until the $k$ nearest neighbors are found.
    \item \textbf{Breadth-First $k$-NN Search}: Performs a breadth-first traversal of the cluster tree,
    pruning clusters using a modified version of the QuickSelect algorithm~\cite{hoare1961algorithm} at each level.
    \item \textbf{Depth-First $k$-NN Search}: Performs a depth-first traversal of the cluster tree and
    uses two priority queues to track candidate clusters and hits.
\end{itemize}

\section{Datasets And Benchmarking}
\label{sec:datasets-and-benchmarks}

We benchmarked panCAKES on a variety of datasets, including RNA sequences, protein sequences data, and set datasets.
We describe each dataset and their associated distance functions.

\subsection{SILVA 18S}
\label{sec:datasets-and-benchmarks:silva-18s}

The SILVA 18S ribosomal RNA dataset~\cite{quast2012silva} contains ribosomal RNA sequences of 2,224,640 genomes in a multiple sequence alignment (MSA)~\cite{thompson2011MSA}.
The sequences are in an MSA that is 50,000 characters wide, most of which are gaps or padding characters for each sequence.
We held out a set of one thousand random sequences prior to clustering to use them as queries for benchmarking.

We built the tree using Hamming distance on the pre-aligned sequences.
For compression when encoding one sequence in terms of another, we store only the indices at which the two sequences differ and the characters at those indices.
For search using the held-out sequences as queries, we simulate having queries that were never in the MSA.
We do this by removing all gaps and padding characters from the queries to get the unaligned sequence.
Then, we use Levenshtein distance~\cite{levenshtein1966binary} to compare the unaligned query to the unaligned versions of the sequences in the tree.

\subsection{GreenGenes}
\label{sec:datasets-and-benchmarks:greengenes}

We use two versions of the GreenGenes dataset~\cite{desantis2006greengenes} for benchmarking: GreenGenes 12.10 and GreenGenes 13.8.

\subsubsection{GreenGenes 12.10}
\label{sec:datasets-and-benchmarks:greengenes:12.10}

This dataset contains 1,075,170 bacterial 16S sequences in an MSA with 7,682 width.
As with SILVA, we held out a set of one thousand random sequences prior to clustering to use them as queries for benchmarking.
We performed clustering, compression, and search in the same way as for the SILVA dataset.

\subsubsection{GreenGenes 13.8}
\label{sec:datasets-and-benchmarks:greengenes:13.8}

This dataset contains 1,261,986 bacterial 16S sequences.
We use the unaligned version of this dataset.
The sequences range in length from 1,111 characters to 2,368 characters.
We held out a set of one thousand random sequences prior to clustering to use them as queries for benchmarking.

We performed clustering and search using Levenshtein distance.
For compression, we used the Needleman-Wunsch algorithm to determine the sequence of edits needed to transform one sequence into another.

\subsection{PDB-seq}
\label{sec:datasets-and-benchmarks:pdb-seq}

This dataset was derived from the protein data bank (PDB)~\cite{bank1971protein}.
We use a subset of the nucleic-acid sequences of all proteins for which structure has been determined.
We have omitted sequences shorter than 30 amino acids, as these are considered polypeptide fragments, and sequences longer than 1000 amino acids, as these represent a small number of outliers such as the protein Titin~\cite{alberts1994molecular}.

We held out a set of one thousand random sequences prior to clustering to use them as queries for benchmarking.
We performed clustering, compression, and search in the same way as for the GreenGenes 13.8 dataset.

\subsection{Kosarak}
\label{sec:datasets-and-benchmarks:kosarak}

Kosarak~\cite{Bodon2003AFA} contains anonymized click-stream data from a Hungarian on-line news portal.
This dataset contains 74,962 sets, with each set containing some combination of members from 27,983 distinct members.
An additional 500 such sets are provided as the query set.
This dataset is available for download from the ANN benchmarks suite~\cite{Bernhardsson2019Benchmarks}.

For clustering and search, we used Jaccard distance, while for the compression, we store only the set differences between pairs of sets.

\subsection{MovieLens-10M}
\label{sec:datasets-and-benchmarks:movielens}

The full dataset~\cite{maxwell2015k} contains 10,000,054 ratings and 95,580 tags applied to 10,681 movies by 71,567 users of the online movie recommender service MovieLens.
The data were filtered (for the ANN-benchmarks suite) down to 69,363 sets with 65,134 total distinct members.
An additional 500 such sets are provided as the query set.

Our approach to clustering, compression and search is identical to that for the Kosarak dataset.

\section{Results}
\label{sec:results}

\subsection{Scaling Behavior of Cluster Radii}
\label{subsec:results:scaling-behavior-of-cluster-radii}

Though it would be ideal if cluster radii decrease with each application of Partition (refer to Algorithm \ref{alg:methods:partition}), this is not the case.
Fortunately, we can make some guarantees about the scaling behavior of cluster radii.

To do this, we must first introduce the concept of \emph{fractal dimensionality} of a dataset.
We assume the manifold hypothesis (refer to~\cite{fefferman2016testing} or Section~\ref{sec:methods:building-the-tree});
in other words, we assume that our dataset is embedded in a $D$-dimensional space, but that the data only occupy a $d$-dimensional manifold, where $d \ll D$. We refer to $d$ as the fractal dimensionality of the dataset.

Though fractal dimensionality is a global property of a dataset, since manifold testing is still extremely computationally costly~\cite{fefferman2016testing}, we infer it from the \emph{local fractal dimensionality} (LFD) at the majority of points in the dataset.

As in~\cite{prior2023cakes}, we define LFD at some length scale around a point in the dataset as:

\begin{equation}
    \frac{\text{log} \left( \frac{|B_X(q, r_1)|}{|B_X(q, r_2)|} \right) }{\text{log} \left( \frac{r_1}{r_2} \right) }
    \label{eq:results:lfd-original}
\end{equation}
where $B_X(q, r)$ is the set of points contained in the metric ball of radius $r$ centered at a point $q$ in the dataset $\textbf{X}$.
Intuitively, LFD measures the rate of change in the number of points in a ball of radius $r$ around a point $q$ as $r$ increases.
When the vast majority of points in the dataset have low ($\ll D$) LFD, we can overload terminology to say that the dataset has low LFD, or simply that the dataset has low fractal dimensionality.

We stress that this concept differs from the \textit{embedding dimension} of a dataset.
To illustrate the difference, consider the SILVA 18S rRNA dataset which contains genomes with unaligned lengths of up to 3,718 base pairs and aligned length of 50,000 base pairs.
The \textit{embedding dimension} of this dataset is at least 3,718 and at most 50,000.
However, due to physical constraints (namely, biological evolution and biochemistry), the data are constrained to a lower-dimensional manifold within this embedding space.
LFD is an approximation of the dimensionality of that lower-dimensional manifold in the ``vicinity'' of a given point.

We prove, in this section, that cluster radii are guaranteed to have decreased after at most $d$ partitions, where $d$ is the fractal dimensionality of the dataset.
First, note that we can describe a $d$-dimensional distribution of data by choosing some set of $d$ mutually orthogonal axes.
Let $2R$ be the maximum pairwise distance among the points in the dataset.
We choose the axes such that the two points that are $2R$ apart lie along one of the axes.
Thus, a $d$-dimensional hyper-sphere of radius $R$ would bound the dataset.
In the worst case, (i.e., with a uniform-density distribution that fills the $d$-sphere), our axes will be such that $2R$ is the maximum pairwise distance \textit{along every axis}.
Such a distribution would also produce a balanced cluster tree.

In this case, Algorithm~\ref{alg:methods:partition} (Partition) will select a maximally distant pair of points to use as poles;
without loss of generality, say we select two points at the extrema one of the $d$ axes.
After one partition, the maximum pairwise distance along that axis will have been halved, i.e., bounded above by $R$.
The next recursive partition will select another of the $d$ axes, and once again, the distance along that axis will be bounded above by $R$.

\begin{figure}[ht!]
    \centering
    \includegraphics[width=3.4in]{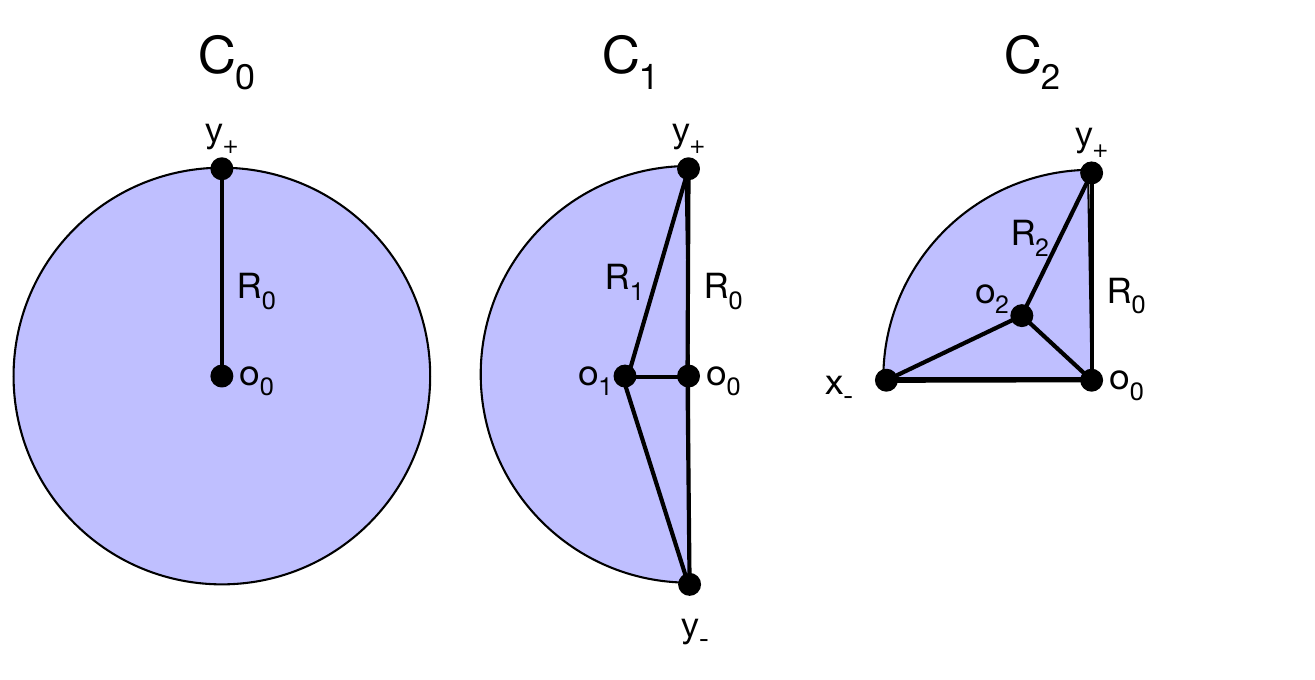}
    \caption{
        Scaling behavior of radii on a two dimensional disk of uniformly distributed points representing the worst-case scenario for a two-dimensional distribution.
        The root cluster $C_0$ has a center $o_0$ and radius radius $R_0$.
        After one application of Partition (Algorithm~\ref{alg:methods:partition}), we have a child cluster $C_1$, with radius $R_1$ and center $o_1$.
        The right triangle formed by $o_0$, $o_1$, and $y_+$ in $C_1$ shows that $R_0 < R_1$.
        Hence, the radius of a child cluster can be larger than that of its parent.
        However, after another application of Algorithm~\ref{alg:methods:partition}, we have consumed both orthogonal axes, as shown in $C_2$.
        Now, clearly $R_2 < R_0$.
    }
    \label{fig:results:radii-scaling-behavior}
\end{figure}

Thus, after at most $d$ partitions, the maximum pairwise distance along each axis will be bounded above by $R$.
The overall (i.e., not restricted to one axis) maximum pairwise distance will be bounded above by $R\sqrt{2}$ for two instances that lie at the extrema of different axes.
See Figure~\ref{fig:results:radii-scaling-behavior} for an example.

After another application of Algorithm~\ref{alg:methods:partition}, we have consumed all the $d$ orthogonal axes, so we cannot have points on the extrema of different axes.
Thus, starting with a cluster $C$ of radius $R$, after at most $d$ partitions, the descendants of $C$ will each have radii bounded above by $\frac{R}{\sqrt{2}}$.
In other words, cluster radii are guaranteed to decrease by a multiplicative factor of $\frac{\sqrt{2}}{2}$ after at most $d$ partitions where $d$ is the local fractal dimension.

Note that, in practice, we never see a balanced cluster tree.
Algorithm~\ref{alg:methods:partition} produces unbalanced trees due to the varying density of the sampling in different regions of the manifold and the low fractal dimension of the data.
Further, in practice, the cluster radii decrease by a factor much larger than $\frac{\sqrt{2}}{2}$ with almost every partition;
the upper bound of $d$ partitions is seldom realized.

\subsection{Compression Upper Bound}
\label{subsec:results:compression-upper-bound}

In this section, we provide an upper bound on the cost of compression in order to illustrate how the cost changes with respect to LFD and depth in the tree.
As in Section~\ref{subsec:results:scaling-behavior-of-cluster-radii}, we assume the worst-case distribution of data and that the local fractal dimension is uniform.

To illustrate the relationship between the LFD and the cost of compression, we introduce the notion of a ``stride'' in the tree.
If $L$ is the LFD of the dataset, we can think of a stride as an $\lceil L \rceil$-depth moving window in the tree.
For example, the first stride includes all the clusters at depth 1 (root only) through depth $\lceil L \rceil$.
The second stride includes the clusters at depths $\lceil L \rceil$ through $2\lceil L \rceil$.
Hereafter, the ceiling is implied when LFD is non-integral; we use simply $L$ to mean $\lceil L \rceil$.

\begin{figure}[ht!]
    \centering
    \includegraphics[width=3.4in]{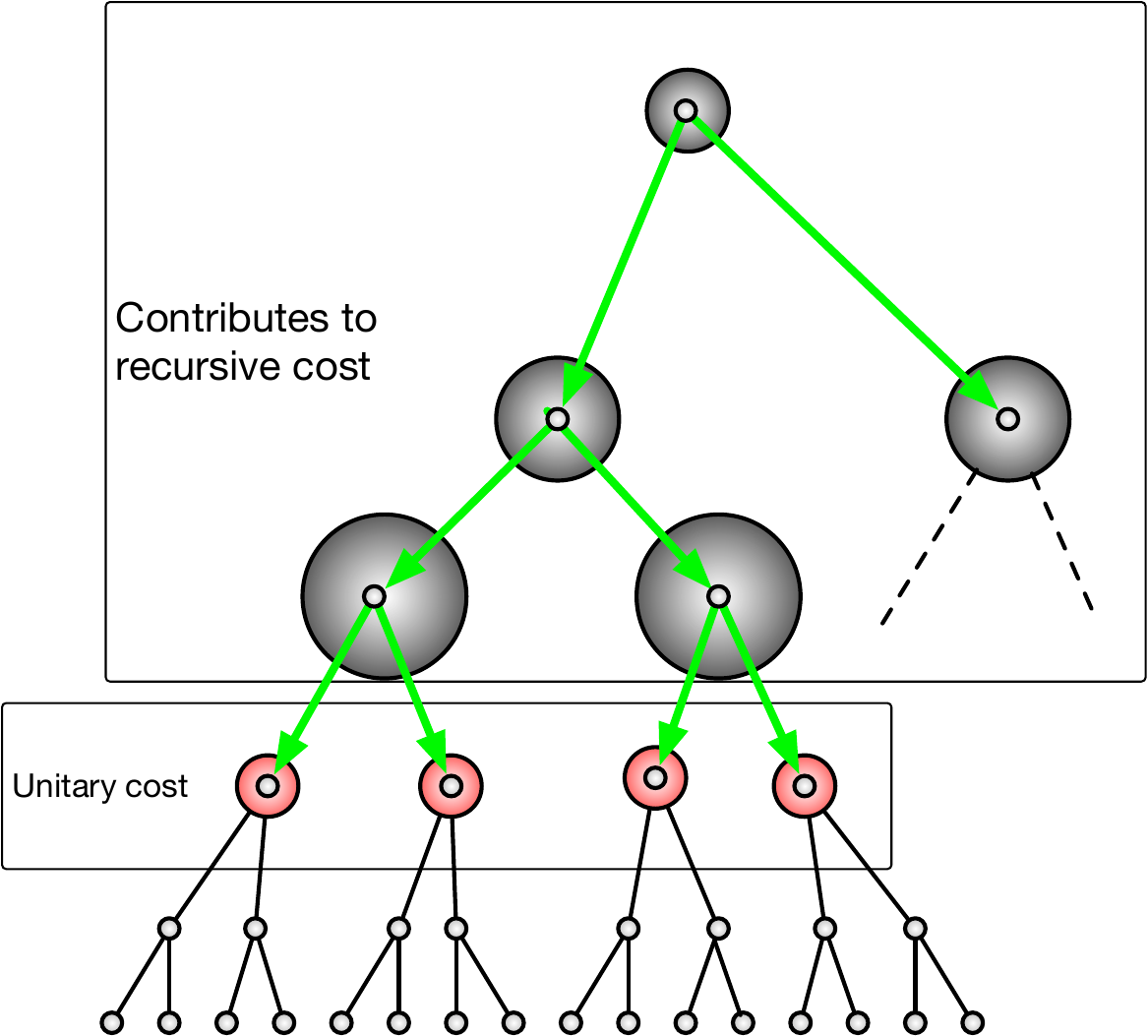}
    \caption{
        Compression cost breakdown assuming LFD = 3 and stride = 1.
        The leaf clusters within the stride (i.e., clusters at depth 4), shaded red, are compressed unitarily. 
        Their ancestors, connected by green edges, are compressed using recursively.
        The recursive cost $T_R$ is the sum of the costs of encoding the green edges.
        The unitary cost $T_U$ is the sum of the costs of encoding all non-center points in the red clusters.
    }
    \label{fig:results:compression-tree}
\end{figure}

To evaluate the cost of compression at stride $i$, we consider the subtree from the root to depth $L\cdot i$.
By Algorithm~\ref{alg:methods:compress}, we would use unitary compression for the leaves of this subtree and recursive compression for all ancestors of those leaves.
Thus, we can think of the cost in terms a unitary component and a recursive component.
We let $T_R$ denote the contribution to compression cost from recursive compression, and let $T_U$ denote the contribution from unitary compression.
Then we have that the total cost $T= T_R + T_U$.
Refer to Figure~\ref{fig:results:compression-tree} for an illustration of this cost breakdown.

To compute $T_R$, we compute the recursive cost within each stride as a function of the stride index $i$, and then sum up over the number of strides $S$.
Each edge (to a child cluster from its parent) in stride $i$ represents one encoding between a pair of cluster centers.
By our analysis in Section~\ref{subsec:results:scaling-behavior-of-cluster-radii}, and by our choice of a distance function under which distance values are proportional to the memory cost of the encoding,
we can use $r_i$, the radius of the shallowest cluster in stride $i$, as an upper bound for the memory cost of each such encoding.

To count the number of edges in a stride, we consider the ``subtrees'' in a stride, where each cluster at the shallowest depth within a stride is the root of a subtree in that stride.
The number of edges in stride $i$ is the product of the number of subtrees at stride $i$ and the number of edges in each subtree.
Let $s_i$ and $e_i$ denote these two quantities respectively.
Then we have that $T_R = \sum_{i = 1}^S r_i \cdot e_i \cdot s_i$.

Then, by the discussion above, we have that
\begin{equation}
    \label{eq:results:r_i}
    r_i = \frac{r}{\sqrt{2}^{(s-1)L}}.
\end{equation}

The number of subtrees at stride $i$ is
\begin{equation}
    \label{eq:results:s_i}
    s_i = 2^{(i-1)L}.
\end{equation}

Since each child cluster has an edge coming to it from its parent, the number of edges in a single subtree at stride $i$ is given by
\begin{equation}
    \label{eq:results:e_i}
    e_i = \sum_{j=2}^L 2^j = 2^{L+1} - 2 = 2 \cdot (2^L - 1).
\end{equation}

Using Equations~\ref{eq:results:r_i}, \ref{eq:results:s_i}, and \ref{eq:results:e_i}, we can compute $T_R$ as follows:
\begin{align}
    T_R &= \sum_{i = 1}^S r_i \cdot e_i \cdot s_i \notag\\
      &= \sum_{i = 1}^S \frac{r}{\sqrt{2}^{(i-1)L}} \cdot \left( 2^{L+1} - 2 \right) \cdot 2^{(i-1)L} \notag \\
      &= 2 \cdot r \cdot (2^L - 1) \cdot \sum_{i = 1}^S 2^{(i-1)\frac{L}{2}} \notag \\
      \label{eq:t_r}
      &= 2 \cdot r \cdot (2^L - 1) \cdot \left( \frac{2^{S\frac{L}{2}} - 1}{2^{\frac{L}{2}} - 1} \right).
\end{align}

To compute $T_U$, we compute the unitary cost of compressing a leaf at stride $S$ and then sum up over the number of leaves at stride $S$.
Let ${\ell}_S$ and $u_S$ denote the number of leaves and the unitary cost of compressing a leaf at stride $S$ respectively. Then we have that $T_U = \ell_S \cdot u_S$.

The number of leaves at stride $S$ is given by
\begin{equation}
    \label{eq:l_s}
    {\ell}_S = 2^{S \cdot L}
\end{equation}

and the unitary cost of compressing a leaf at stride $S$ is given by
\begin{equation}
    \label{eq:u_s}
    u_S = \frac{r}{\sqrt{2}^{S \cdot L}} \cdot \left( \frac{|C|}{2^{S\cdot L}} - 1\right).
\end{equation}

The first factor in Equation~\ref{eq:u_s} uses the radii scaling behavior guarantee from Section~\ref{subsec:results:scaling-behavior-of-cluster-radii} and the second factor is the number of non-center points in leaves at depth $S\cdot L$ (recall that we are assuming a balanced clustering).

By substituting Equations~\ref{eq:l_s} and~\ref{eq:u_s} into the definition of $T_U$, we have that
\begin{equation}
    \label{eq:t_u}
    T_U = 2^{S \cdot L} \cdot \frac{r}{\sqrt{2}^{S \cdot L}} \cdot \left( \frac{|C|}{2^{S\cdot L}} - 1\right).
\end{equation}

Finally, combining Equations~\ref{eq:t_r} and \ref{eq:t_u}, we have that the total cost of compression is given by
\begin{gather}
    \label{eq:total-cost}
    T =
    \underbrace{
        2 \cdot r \cdot (2^L - 1) \cdot \frac{2^{S\frac{L}{2}} - 1}{2^{\frac{L}{2} - 1}}
    }_{\textrm{recursive}}
    \ + \
    \underbrace{
        2^{S \cdot L} \cdot \frac{r}{\sqrt{2}^{S \cdot L}} \cdot \left( \frac{|C|}{2^{S\cdot L}} - 1\right)
    }_{\textrm{unitary}}.
\end{gather}

Equation \ref{eq:total-cost} shows how the cost scales with the number of strides $S$ and the local fractal dimension $L$.
As the number of strides into the tree increases, the recursive cost $T_R$ increases exponentially, while the unitary cost $T_U$ decreases exponentially.
Similarly, as the local fractal dimension $L$ increases, the recursive cost increases exponentially, while the unitary component $T_U$ decreases exponentially.

\subsection{Compression Ratios}
\label{sec:results::compression-multipliers}

Table~\ref{tab:compression-ratios} reports the compression ratios of panCAKES and gzip.
We observe that panCAKES offers significantly better compression than gzip for the SILVA 18-S dataset, with a compression ratio of 69.96x for panCAKES compared with 4.397 for gzip.
We observe that panCAKES offers comparable compression to gzip for the set datasets, Kosarak and MovieLens 10M.
For the GreenGenes 13.5 and 12.10 datasets, panCAKES offers worse compression than gzip.
For the PDB-seq protein dataset, panCAKES does not offer any compression at all, as illustrated by the compression ratio of 0.61x.

\begin{table*}[!t]
    \caption{Compression Ratios.}
    \label{tab:compression-ratios}
    \begin{center}
        \begin{sc}
            \begin{tabular}{|c|c|c|c|c|c|}
                \hline
                \multirow{2}{*}{\textbf{Dataset}} & \textbf{Raw Data} & \textbf{Gzip} & \textbf{Gzip} & \textbf{panCAKES} & \textbf{panCAKES}   \\
                & \textbf{(MB)} & \textbf{(MB)} & \textbf{Ratio} & \textbf{(Data + Tree) (MB)} & \textbf{Ratio} \\
                \hline
                GreenGenes 13.5     & 1,740            & 256            & 6.80x               & 631 + 147                         & 2.24x                      \\
                \hline
                GreenGenes 12.10    & 7,887            & 146            & 54.02x              & 289 + 97                          & 19.97x                     \\
                \hline
                Silva 18-S          & 107,676          & 4,397          & 24.49x              & 1,320 + 219                       & 69.96x                      \\
                \hline
                PDB-seq             & 251              & 42             & 5.98x               & 326 + 88                          & 0.61x                       \\
                \hline
                Kosarak             & 33               & 11             & 3.00x               & 10 + 1.5                          & 2.87x                       \\
                \hline
                MovieLens 10M       & 63               & 19             & 3.32x               & 18 + 1.3                          & 3.26x                        \\
                \hline
            \end{tabular}
        \end{sc}
    \end{center}
    \vskip -0.1in
\end{table*}

\subsection{Search}
\label{subsec:results:search}

Table~\ref{tab:search-time} reports the search time in seconds per query for our four algorithms on \emph{raw datasets}.
Table~\ref{tab:compressive-search-time} reports the same figure for datasets compressed using panCAKES.
In Table~\ref{tab:summary-table}, we report the ratio of compressive search time to raw data search time for each dataset.
We observe that, with the exception of the PDB-Seq dataset, panCAKES loses search speed while gaining compression.
For $\rho$-NN search, the loss in search speed ranges from 1.28x to 28.26x.
For repeated $\rho$-NN search, figure ranges from 1.20x to 8.99x.
BFS-$k$-NN search loses speed by a factor of 0.90x to 6.50x, and DFS-$k$-NN search loses speed by a factor of 0.27x to 4.18x.

The $k$-NN search algorithm using repeated $\rho$-NN, as described in~\cite{prior2023cakes}, is a special case here.
It relies on the ability to find incrementally more points with small increases in the search radius for $\rho$-NN.
This works well under Levenshtein edit distance but breaks down under Jaccard distance and is unable to sufficiently increase the radius to find additional potential hits.
So, we omit this algorithm from the benchmarks on the set datasets.

\begin{table*}[!t]
    \caption{Search time in seconds per query.}
    \label{tab:search-time}
    \begin{center}
        \begin{sc}
            \begin{tabular}{|c|c|c|c|c|}
                \hline
                \textbf{Dataset}  & \textbf{$\rho$-NN} &  \textbf{Repeated $\rho$-NN} & \textbf{BFS $k$-NN} & \textbf{DFS $k$-NN}   \\
                \hline
                GreenGenes 13.5       & 2.53              & 9.64                          & 11.71               & 11.96                \\
                \hline
                GreenGenes 12.10      & 3.17              & 21.52                          & 24.43                & 18.72                \\
                \hline
                Silva 18-S            & 5.87              & 43.48                         & 35.59               & 154.66                 \\
                \hline
                PDB-seq               & 0.67               & 2.07                          & 0.91                 & 2.63                 \\
                \hline
                Kosarak               & 6.20 $\times 10^{-3}$  & -                         & 2.56 $\times 10^{-3}$  & 1.75 $\times 10^{-3}$ \\
                \hline
                MovieLens 10M         & 7.38 $\times 10^{-3}$ & -                         & 2.83 $\times 10^{-2}$ & 2.60 $\times 10^{-2}$ \\
                \hline
            \end{tabular}
        \end{sc}
    \end{center}
    \vskip -0.1in
\end{table*}

\begin{table*}[!t]
    \caption{Compressive Search time in seconds per query.}
    \label{tab:compressive-search-time}
    \begin{center}
        \begin{sc}
            \begin{tabular}{|c|c|c|c|c|}
                \hline
                \textbf{Dataset}  & \textbf{$\rho$-NN} &  \textbf{Repeated $\rho$-NN} & \textbf{BFS $k$-NN} & \textbf{DFS $k$-NN}   \\
                \hline
                GreenGenes 13.5       & 41.78             & 86.68                        & 76.12                & 50.04                \\
                \hline
                GreenGenes 12.10      & 89.57             & 144.59                       & 155.31               & 102.20              \\
                \hline
                Silva 18-S            & 77.91             & 161.32                       & 166.29               & 130.26                 \\
                \hline
                PDB-seq               & 0.85               & 2.48                         & 0.82                  & 0.72                 \\
                \hline
                Kosarak               & 2.1 $\times 10^{-2}$ & -                          & 2.68 $\times 10^{-2}$ & 2.84 $\times 10^{-2}$ \\
                \hline
                MovieLens 10M         & 3.28 $\times 10^{-2}$ & -                        & 50.0 $\times 10^{-2}$  & 49.4 $\times 10^{-2}$ \\
                \hline
            \end{tabular}
        \end{sc}
    \end{center}
    \vskip -0.1in
\end{table*}

In Table~\ref{tab:summary-table}, we summarize the compression ratios achieved by panCAKES and the trade-off in loss of search speed.

\begin{table*}[!t]
    \caption{Summary Table of the factors by which panCAKES loses search speed while gaining compression.}
    \label{tab:summary-table}
    \begin{center}
        \begin{sc}
            \begin{tabular}{|c|c|c|c|c|c|}
                \hline
                \textbf{Dataset} & \textbf{$\rho$-NN}  & \textbf{Repeated $\rho$-NN}  & \textbf{BFS $k$-NN} &  \textbf{DFS $k$-NN} & \textbf{Compression}   \\
                \hline
                GreenGenes 13.5  & 16.49                     & 8.99                    & 6.50                & 4.18                 & 2.24                  \\
                \hline
                GreenGenes 12.10 & 28.26                     & 6.72                    & 6.36                & 5.46                 & 19.97                 \\
                \hline
                Silva            & 13.29                     & 3.71                    & 4.67                & 0.84                 & 69.96                 \\
                \hline
                PDB-seq          & 1.28                      & 1.20                    & 0.90                & 0.27                 & 0.61                  \\
                \hline
                Kosarak          & 3.39                      & -                       & 1.04                & 1.62                 & 2.87                  \\
                \hline
                MovieLens        & 4.44                      & -                       & 1.77                & 1.90                 & 3.26                  \\
                \hline
            \end{tabular}
        \end{sc}
    \end{center}
    \vskip -0.1in
\end{table*}

\section{Discussion and Future Work}
\label{sec:discussion-and-future-work}

We have presented panCAKES, a novel approach to compression and compressive-search for big data. Our approach allows for efficient similarity search \emph{without decompressing the entire dataset}. We achieve this result by encoding each data point as its set of differences from some representative data point; these differences can then be applied to the representative to reconstruct the original data. This reconstruction need not be applied to the entire dataset (i.e., the root of the CLAM tree), but can instead be applied to any arbitrary subtree corresponding to the result set. The approach is generic over any distance function for which the distance between two points is proportional to the memory cost of storing an encoding of one in terms of the other. In essence,  panCAKES can be thought of as a simultaneous database and compression index, allowing for selective decompression based on a query.

\subsection{Discussion}
\label{subsec:discussion-and-future-work:discussion}

Our results illustrate the tradeoff between compression ratios and search speed.
Though panCAKES had the greatest compression ratios on the two aligned genomic datasets (SILVA and GreenGenes 12.10), these datasets exhibited relatively slow search times.
Notably, panCAKES achieves ~3x better compression on the SILVA dataset than gzip.
While panCAKES compression is comparable to that of gzip on the two set datasets (Kosarak and MovieLens), $k$-NN search is only mildly slower on the compressed representations of these datasets than on the raw data.
We stress that the primary advantage of panCAKES is that search can happen \emph{without} decompressing the whole dataset, which is not possible with general purpose compression algorithms such as gzip. This essential when the dataset is vastly larger than what can fit in system memory.

We expect that the compression ratio is highly dependent on the amount of self-similarity present in the dataset.
This is supported by the fact that SILVA-18S, which is known to have many nearly redundant sequences~\cite{quast2012silva}, exhibits the highest compression ratio of all datasets.

The protein data bank sequences (PDB) form an unusual case, in that PDB exhibits selection bias;
it is unusual for protein structures to be deposited in the PDB when they exhibit high sequence similarity to existing entries~\cite{daniels2011touring}.
Thus, we see that the advantages of self-similar (and thus, compressible) sequences are elusive here.
We expect that should protein space be better explored, we would see results in line with what we observe for genomic data.

\subsection{Future Work}
\label{subsec:discussion-and-future-work:future-work}

As future work, we intend to benchmark panCAKES on additional types of data satisfying the properties outlined in Section~\ref{sec:introduction}.
These include integer vector data under Manhattan distance, image datasets under Wasserstein distance, and radio frequency data under Dynamic Time Warping.

We would also like to better understand the relationship between panCAKES compression ratio and other intrinsic properties of a dataset, such as fractal dimension and metric entropy~\cite{yu2015entropy}.

With strings datasets, it would also be valuable to explore an alternative approach to storing the edits from Needleman-Wunsch.
When the sequences are very long, the memory cost of storing the index of an edit can be high.
For certain dimensionalities, we may be able to store edits in a more memory-efficient way by instead storing only the differences between the indices of consecutive pairs of edits.
Intuitively, when the gap between successive edits is large relative to the length of the sequence, this implies that there few edits, and hence the total memory cost would be low.
Alternatively, if there are many edits, the distance between successive edits would be small and so each difference will require fewer bits to store, thus decreasing the memory cost.

As an other extension, we plan to adapt our approach to offer lossy compression for datasets of floating point vectors.
We also intend to explore how lossy compression could allow for faster, exact search.
For example, suppose we compress a dataset using panCAKES, maintain a decompressed version of the dataset on disk, and keep a one-to-one mapping between the compressed and decompressed points.
We can then perform similarity search in the compressed space.
Using the one-to-one mapping, for each point in the results set, one could determine the corresponding decompressed point on disk.
Using looser search criteria (values of $\rho$ or $k$) may mitigate the loss of search accuracy due to the lossy compression.

Additionally, we intend to explore improvements to the compression algorithm in Section~\ref{sec:methods:compression}.
Algorithm~\ref{alg:methods:compress} trims the tree such that only the shallowest clusters marked for unitary compression are kept;
any descendants of these clusters are discarded.
This approach is greedy and may not be optimal;
we may achieve better compression ratios by deferring the trimming of descendants until reaching a deeper depth in the tree.

\ifarxiv
\else
\bibliographystyle{IEEEtran}
\fi
\bibliography{references}

\end{document}